\documentclass[twocolumn,showpacs,preprintnumbers,amsmath,amssymb]{revtex4}

\usepackage{graphicx}
\usepackage{dcolumn}
\usepackage{bm}

\begin{document}


\title{Observation of Four-Photon de Broglie Wavelength\\by State Projection Measurement}

\author{F. W. Sun$^1$, B. H. Liu,$^{1}$ Y. F. Huang,$^{1}$, Z. Y. Ou$^{1,2}\footnote{E-mail:
zou@iupui.edu}$, and G. C. Guo$^1$}
 \affiliation{$^1$Key Laboratory of Quantum Information,
 University of Science and Technology of China, \\CAS, Hefei, 230026, the People's Republic of China
 \\$^2$Department of Physics, Indiana
University-Purdue University Indianapolis, 402 N. Blackford
Street, Indianapolis, IN 46202}

\date{\today}

\begin{abstract}
A measurement process is constructed to project an arbitrary
two-mode $N$-photon state to a maximally entangled $N$-photon
state (the {\it NOON}-state). The result of this projection
measurement shows a typical interference fringe with an $N$-photon
de Broglie wavelength. For an experimental demonstration, this
measurement process is applied to a four-photon superposition
state from two perpendicularly oriented type-I parametric
down-conversion processes. Generalization to arbitrary $N$-photon
states projection measurement can be easily made and may have wide
applications in quantum information. As an example, we formulate
it for precision phase measurement.
\end{abstract}

\maketitle

It has now been well established that only nonclassical states of
light such as squeezed states can have accuracy in precision phase
measurement better than the standard quantum limit which goes as
$1/\sqrt{N}$ for an average photon number of $N$ of the phase
probing field \cite{cav,xia,gra}. It was argued \cite{ou} that
there exists an ultimate limit, i.e., the Heisenberg limit
\cite{hei} in precision phase measurement for an arbitrary state
of the probe light. The Heisenberg limit goes as $1/N$ for an
average photon number of $N$. A number of quantum states
\cite{ou,bon,yur,hol,jac,bol} have been identified that can
achieve such a limit when used to probe a phase shift. Among them,
the {\it NOON}-state of a two-mode maximally entangled $N$-photon
superposition state has been in the forefront of discussions
recently \cite{bot,kok,lee,kok2,wal,mit,lei,roo}. Such a state is
described as $ |NOON\rangle = (|N,0\rangle -
|0,N\rangle)\sqrt{2}$. An $N$-photon coincidence measurement in
the superposition of the two modes gives a dependence of $1-\cos
N\varphi$ on the single photon phase shift $\varphi$. This phase
dependence is typical of a fringe pattern with an $N$-photon de
Broglie wavelength and can be managed to achieve the Heisenberg
limit in accuracy in the measurement of the single-photon phase
shift $\varphi$.

Because of the difficulty in preparing multi-particle
entanglement, only up to four-photon de Broglie wavelength has
been demonstrated so far \cite{wal,mit} through some special
interference effect to cancel the unwanted states of
$|N-1,1\rangle, |N-2,2\rangle, ...$ etc.  A number of schemes
\cite{hof,sha,wan} have been proposed. Among them, the proposal by
Hofmann \cite{hof} can be generalized to an arbitrary $N$ and was
demonstrated recently by Mitchell {\it et al.} \cite{mit} for
$N=3$ case. It is worth noting that most of the schemes are based
on an $N$-photon coincidence measurement to discriminate against
the states of photon number $<N$.

In this letter, we will approach this problem from a completely
different direction, namely, the measurement process. We will
describe an interference scheme similar to that of Hofmann
\cite{hof} to form an $NOON$ state projection measurement. The
scheme only depends on the contribution from $NOON$ state while
discarding all other orthogonal states in an arbitrary $N$-photon
state. We demonstrate our projection method experimentally with a
four-photon superposition state from two perpendicularly oriented
type-I parametric down-conversion processes. Although the quantum
state is not a $NOON$ state, the projection measurement allows us
to demonstrate an interference fringe pattern with the typical
four-photon de Broglie wavelength. Furthermore, our projection
method can be easily generalized to an arbitrary $N$-photon state.

\begin{figure}[htb]
\begin{center}
\includegraphics[width= 3.3in]{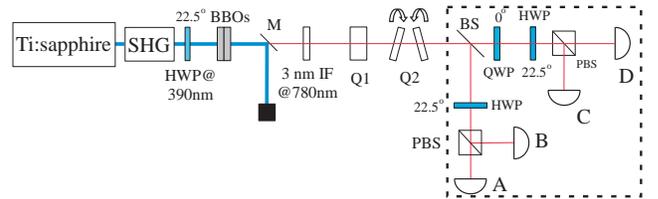}
\end{center}
\caption{\em Layout for four-photon NOON state projection
measurement with parametric down conversion. } \label{fig1}
\end{figure}

The four-photon NOON state projection measurement is depicted in
the detection part of Fig.1 (inside the dotted box). The operators
of the four detectors are related to the horizontal and vertical
components of the input field as $ \hat b_A= (\hat a_H-\hat a_V)/2
+ \hat b_{A0},\hat b_B= (\hat a_H+\hat a_V)/2+ \hat b_{B0}, \hat
b_C= (\hat a_H-i\hat a_V)/2+ \hat b_{C0}, \hat b_D= (\hat
a_H+i\hat a_V)/2+ \hat b_{D0} $ Here $\hat b_{n0}~(n=A,B,C,D)$ are
some operators related to the vacuum modes $\hat a_{0H,V}$ in the
unused beam splitter input port and make no contribution to photon
detection. The four-photon coincidence rate of detectors $A,B,C,D$
is then proportional to
\begin{eqnarray}
&&P_4 = \langle \Phi_4|\hat b_D^{\dagger}\hat b_C^{\dagger}\hat
b_B^{\dagger}\hat b_A^{\dagger}\hat b_A \hat b_B\hat b_C\hat
b_D|\Phi_4\rangle \cr &&\hskip 0.2 in =   |\langle 0|\hat
a_H^4-\hat a_V^4|\Phi_4\rangle|^2/2^{8}={3\over 16} |_4\langle
NOON|\Phi_4\rangle|^2.~~~~~
\end{eqnarray}
Here $|\Phi_4\rangle=\sum_k c_k|N-k,k\rangle$ is a four-photon
state.

To experimentally test the projection measurement scheme, we apply
it to a quantum state produced from two identical but
perpendicularly oriented Type-I parametric down-conversion
processes. The Hamiltonian for the production of such a state has
the form of
\begin{eqnarray}
\hat H = i\hbar \chi (\hat a_{H}^{\dagger 2} + \hat a_{V}^{\dagger
2} ) + H.c.\label{H}
\end{eqnarray}
The four-photon part of the quantum state for weak interaction
thus has the form of
\begin{eqnarray}
&&|\Phi_4\rangle = {\eta^2 \over 2}(\hat a_{H}^{\dagger 2} + \hat
a_{V}^{\dagger 2})^2|vac\rangle \cr &&\hskip 0.3in\propto
\sqrt{3\over 8}\Big(|4,0\rangle+|0,4\rangle \Big) + {1\over
2}|2,2\rangle.\label{state}
\end{eqnarray}
The $NOON$ state projection gives
\begin{eqnarray}
P_4 \propto  1 - \cos 4\varphi,\label{P4}
\end{eqnarray}
where $\varphi$ is the single photon phase difference between the
H and V polarizations. This shows the typical fringe pattern with
a four-photon de Broglie wavelength.

\begin{figure}[htb]
\begin{center}
\includegraphics[width= 3.0in]{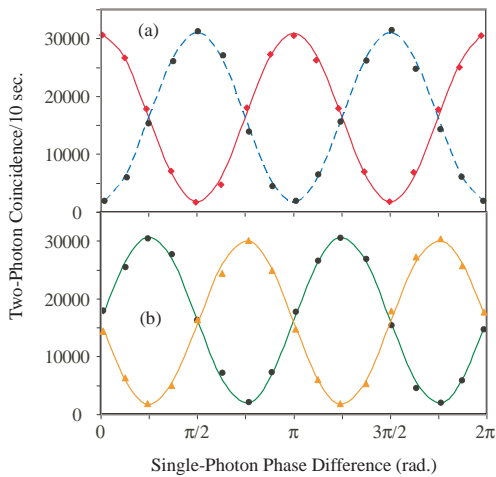}
\end{center}
\caption{\em Two-photon coincidence as a function of single-photon
phase difference $\varphi$. (a) Coincidences between AB (circle)
and between CD (diamond). The continuous curves are the
least-square fit to the function $N_{2c0}(1\pm v_2 \cos
2\varphi)$. (b) Coincidences between AD (circle) and between AC
(triangle). The continuous curves are the least-square fit to the
function $N_{2c0}(1\pm v_2 \sin 2\varphi)$.} \label{fig2}
\end{figure}

Experimental arrangement is sketched in Fig.1. A Coherent Mira
Ti:sapphire laser with 150 fs pulse width and 76 MHz repetition
rate is frequency-doubled to 390 nm. The harmonic field of 200 mW
serves as the pump field for parametric down-conversion in two
2-mm thick BBO crystals cut for Type-I process. The polarization
of the pump field is rotated by a half wave plate (HWP) to
45$^{\circ}$ and the two crystals are oriented so that their fast
(or slow) axes are perpendicular to each other. Thus the
horizontal polarization of the pump field is the pump for one
crystal and the vertical polarization for another crystal. The
quantum state from the two crystals has the form of
Eq.(\ref{state}). The down-converted light first passes through an
interference filter centered at 780 nm with a bandwidth of 3 nm.
Then stacks of quartz plates ($Q1$) are used to compensate the
delay between H and V polarizations of the down-converted photons.
Another set ($Q2$) of quartz plates are used for precision phase
control between H and V polarizations. The compensated light is
directed to the $NOON$ state projection measurement assembly.
Fig.2a shows the two-photon coincidence $N_c$ in 10 seconds
between detectors A and B and between detectors C and D as a
function of single-photon phase difference $\varphi$. Fig.2b
presents data for coincidence measurement between A and C and
between A and D. Coincidence data between B and D is nearly
identical to that between A and C and coincidence between C and B
is same as that between A and D. These two are not plotted.
$N_c^{(AB)}$ and $N_c^{(CD)}$ are least-square-fitted to
$N_{c0}(1\pm v_2\cos2\varphi)$ and $N_c^{(AC)}$ and $N_c^{(AD)}$
to $N_{c0}(1\pm v_2\sin 2\varphi)$, respectively. They have an
average visibility of $v_2 = 0.88$ after background subtraction.
The four-photon coincidence counts in 5 minutes are plotted in
Fig.3(a) as a function of single-photon phase difference. The data
is after background subtraction. The continuous curve is a
least-square fit of the discrete experimental data to
$N_{4c0}(1-V_4\cos 4\varphi)$ with a visibility of $V_4=0.57\pm
0.05$.

\begin{figure}[htb]
\begin{center}
\includegraphics[width= 2.5in]{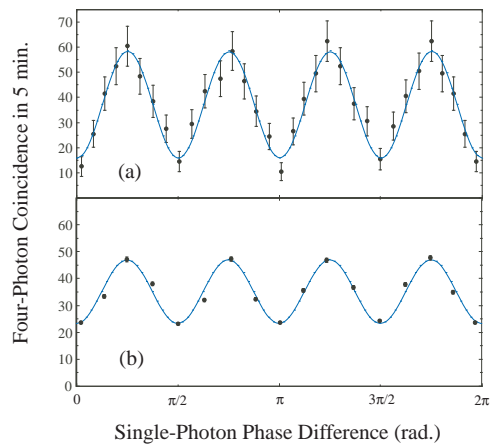}
\end{center}
\caption{\em Four-photon coincidence as a function of
single-photon phase difference $\varphi$. The solid curves are a
least square fit to the function $N_{4c0}(1-V_4 \cos 4\varphi)$.
(a) Measured four-photon coincidence with $V_4 = 0.57\pm 0.05$ and
$N_{4c0} = 37\pm 1$. (b) Indirectly measured four-photon
coincidence for $2\times 2$ case with $V_4^{(0)} = 0.34\pm 0.01$
and $N_{4c0} = 35\pm 1$. } \label{fig3}
\end{figure}

Two observations can be made from Fig.3(a). Firstly, it indeed
gives the $4\varphi$ dependence on the phase as predicted by
Eq.(\ref{P4}). Secondly, the visibility of the sinusoidal
modulation is only 57\%, far short of the 100\% predicted in
Eq.(\ref{P4}). Part of the explanation of this discrepancy is from
the less than perfect spatial mode match, which already results in
only 88\% visibility in the two-photon interference curves in
Fig.2. This mode match problem arises from the walk-off between
the ordinary and the extra-ordinary rays in the crystals. Because
of this, the spatial mode is elongated in one direction for the
first crystal but in a perpendicular direction for the second
crystal. We may reduce this detrimental effect by using a mildly
focussed pump beam.

However, most of the reduction in visibility actually comes from
temporal mode match \cite{ou2,ou3,tsu,ou4}. It is well known that
the Hamiltonian in Eq.(\ref{H}) produces a pair of entangled
photons in a state of $ |\Phi^+\rangle = (|2H\rangle
+e^{2i\varphi}|2V\rangle/\sqrt{2}$. Four photons can be generated
from two-pair production. But it is not guaranteed that the two
pairs are always generated in the same time. For short pump
duration, the pairs are overlapping and indistinguishable from
each other, this situation is generally known as $4\times 1$ case
and the quantum state is given by Eq.(\ref{state}) with the $NOON$
state projection measurement result given in Eq.(\ref{P4}). On the
other hand, when the pump duration is long, the two pairs are
produced mostly at different times and are independent of each
other. This is the $2\times 2$ case with a quantum state described
by $|\Phi_{2\times 2} \rangle =
|\Phi^+\rangle_1\otimes|\Phi^+\rangle_2$. Here 1 and 2 denote the
two distinguishable pairs. It can be shown \cite{sun} that the
$2\times 2$ case will also show an interference fringe pattern
with four-photon de Broglie wavelength but the interference
visibility is only $V_4^{(0)} = 3/7= 0.43$.

The real system is actually in between the two extreme cases
described above. Tsujino et al. \cite{tsu} described the system as
a statistical mixture of the two cases. A multi-mode approach
\cite{ou3,ou4,sun} provides a complete account of all the mode
match effects discussed here. From there, a general formula for
the four-photon interference visibility \cite{sun} can be derived
and has the form of
\begin{eqnarray}
V_4 = 3({\cal A}+2{\cal E})v_2^2/[(6+v_2^2){\cal A}+2{\cal
E}(3-2v_2)],\label{V}
\end{eqnarray}
where ${\cal A}$ is proportional to the accidental rate of
two-pair production and ${\cal E}(\le {\cal A})$ characterizes the
overlap between the two pairs. When ${\cal E}={\cal A}$, the two
pairs are completely overlapping and the four photons are in an
indistinguishable entangled state described by Eq.(\ref{state}).
This is the $4\times 1$ case. On the other hand, when ${\cal E} =
0$, the two pairs are completely separated from each other and
become independent. This is the $2\times 2$ case. Substituting the
observed values of $V_4 = 0.57\pm 0.05$ and $v_2=0.88$ in
Eq.(\ref{V}) and solving for ${\cal E}/{\cal A}$, we obtain ${\cal
E}/{\cal A} = 0.49\pm 0.12$.

For the situation when the two pairs are independent of each
other, the four-photon coincidence rate $R_{4}^{(0)}$ can be
deduced from the measured two-photon coincidence rates among all
four detectors. In this case, four-photon probability at four
detectors in one pump pulse is simply the sum of all possible
products of the two-photon probabilities at one and other pair of
detectors, that is,
\begin{eqnarray}
P_{4}^{(0)} = P_{AB}P_{CD} + P_{AC}P_{BD} + P_{AD}P_{BC},
\end{eqnarray}
where $P_{ij} (i,j=A,B,C,D)$ is two-photon probability in
detectors $i$ and $j$. In terms of coincidence rate, we have
\begin{eqnarray}
R_4^{(0)} = \Big(R_{AB}R_{CD}+
R_{AC}R_{BD}+R_{AD}R_{BC}\Big)/R,\label{P40}
\end{eqnarray}
where $R = 76$ MHz is the repetition rate of pump pulses. The
two-photon rates were measured and shown in Fig.2 and $R_4^{(0)}$
can then be derived from Eq.(\ref{P40}). Fig.3(b) shows the
derived $R_4^{(0)}$ as a function of single-photon phase
difference. The visibility from a least square fit has the value
of $V_4^{(0)} = 0.34$, which is exactly same as that derived from
Eq.(\ref{V}) when we take ${\cal E}/{\cal A}=0$ and $v_2=0.88$.

The value of ${\cal E}/{\cal A}$ can be independently measured in
our experiment. If we rotate the pump polarization to either $H$
or $V$-direction, only one crystal will have down-conversion and
from Ref.\cite{sun} we have
\begin{eqnarray}
R_4 \propto {\cal A} + 2{\cal E},~~~ {\rm or }~~~R_4 = R_{4}^{(0)}
(1 + 2{\cal E}/{\cal A}),\label{EA}
\end{eqnarray}
where $R_{4}^{(0)}$ is the accidental four-photon coincidence rate
for the $2\times 2$ case (${\cal E} =0$). $R_{4}^{(0)}$ can be
calculated from the measured two-photon coincidence rates among
all four detectors from Eq.(\ref{P40}). Experimentally, when we
set the pump polarization to $H$-direction, we observed
$R_{AB}=777/s, R_{CD}=892/s, R_{AC}=800/s, R_{DB}=862/s,
R_{AD}=823/s, R_{CB}=847/s$, and $R_4 = (103\pm 10)/30 {\rm min}$
after back ground corrections. This gives rise to $R_4^{(0)} =
0.0274/s = 49.3/30{\rm min}$ from Eq.(\ref{P40}). By
Eq.(\ref{EA}), we then have ${\cal E}/{\cal A} = 0.54\pm 0.05$.
Within the error allowance, this value coincides with the value
derived from the visibility consideration.

The value of ${\cal E}/{\cal A}$ gives a measure of how
indistinguishable the two pairs of photons are from each other in
parametric down-conversion. It has a complicated dependence on
temporal/spectral mode structure such as the pump pulse width, the
down-conversion bandwidth, and optical filtering before detection.
It also depends on the spatial mode structure such as the pump
field focusing. Generally speaking, narrowing the pump pulse width
and the crystal length to have a well defined time of pair
production will increase the value of ${\cal E}/{\cal A}$.
Frequency filtering can also enforce a good temporal mode match.
However, these measures will reduce the count rate of the pairs
and make the statistics of the data even poorer than present. For
the spatial mode, we find that ${\cal E}/{\cal A}$ increases when
we have a tight focus of the pump. But this will cause poor
spatial mode match between the two crystals and decrease the
visibility $v_2$ of two-photon interference, leading to a reduced
$V_4$. The observed value of ${\cal E}/{\cal A} = 0.49\pm 0.12$ is
a trade-off among a good counting rate, a relatively high
two-photon visibility, and a moderate four-photon interference
visibility.

As claimed by many \cite{ou,bol,wal,mit,lei}, the observation of
multi-photon de Broglie wavelength may lead to improvement in
precision phase measurement. For the current $NOON$ state
projection measurement, however, the probability $P_N$ of a
successful projection to the $NOON$ state is a very small number
if we do not use a $NOON$ state. When $P_N$ goes to zero for large
$N$, we may not claim the Heisenberg limit even if we observed the
$N$-photon de Broglie wavelength. This is because we need to apply
the scheme $1/P_N$ times before we can have a successful
projection. Thus the total photon number is $N/P_N$. It can be
shown\cite{sun2} that the best phase uncertainty for this case is
\begin{eqnarray}
\Delta\varphi_m = \sqrt{(2-P_N) /P_N}/N.\label{dphi}
\end{eqnarray}
For finite $P_N$, this still gives $\sim 1/N$, i.e., the
Heisenberg limit. But for most state available in laboratory, $P_N
\sim 0$ as $N\rightarrow \infty$. For example, the state in
parametric down-conversion from the Hamiltonian in Eq.(\ref{H}) is
given by
\begin{eqnarray}
|PDC_N\rangle =\sum_{n=0}^{N}{\sqrt{(2N-2n)!(2n)!}\over
2^{N}(N-n)!n!} |2(N-n),2n\rangle,\label{pdc}
\end{eqnarray}
and we can easily find $P_N(PDC) \rightarrow 1/\sqrt{N}$. This
leads to $\Delta\varphi_m \sim N^{-3/4}$ from Eq.(\ref{dphi}).

On the other hand, we may choose a different projection
measurement.  This relies on the generalization of the $NOON$
state projection measurement discussed in current paper to an
arbitrary $N$-photon superposition state of $|\Phi_N\rangle
=\sum_{n=0}^{N}c_n|N-n,n\rangle$. This is straightforward and is
shown in Ref.\cite{sun}.

With the arbitrary state projection measurement, we may choose the
following strategy. For a given $N$-photon state $|\Phi_N\rangle$,
we find its relevant orthogonal state through phase change, that
is, find the minimum $\varphi_m$ so that $\langle \Phi_N |\Phi_N
(\varphi_m)\rangle =0$. For the NOON state, for example, the
orthogonal state is $|NOON+\rangle = (|N,0\rangle + |0,N\rangle)
/\sqrt{2}$ by a phase shift of $\varphi_m=\pi/N$ (from
Eq.(\ref{P4})). Now we prepare the incoming field in the
orthogonal state $|\Phi_N (\varphi_m)\rangle$ and our $|\Phi_N
\rangle$-state projection measurement gives $P_{\Phi_N }(|\Phi_N
(\varphi_m)\rangle) =|\langle \Phi_N|\Phi_N (\varphi_m)\rangle|^2
= 0$. Suppose there is a phase shift of $\delta =-\varphi_m$
between H and V to shift the $|\Phi_N (\varphi_m)\rangle$ state
back to $|\Phi_N\rangle$ so that $P_{\Phi_N}(\delta) =1$. Then by
detecting an $N$-photon coincidence, we detect the small phase
shift $\delta =\varphi_m$. For the NOON state, we have the
Heisenberg limit with $\delta =\varphi_m = \pi/N$.

Of course, the $NOON$ state is not easy to generate in laboratory
but the state in Eq.(\ref{pdc}) is available from parametric
down-conversion. For this state, the minimum phase shift
$\varphi_m$ for its orthogonal state as $\varphi_m = 1.53\pi/N$
for large $N$ \cite{sun2}. Although this scheme is a little worse
than the $NOON$ state case, it indeed approaches the Heisenberg
limit.

In summary, we have demonstrated experimentally the four-photon de
Broglie wavelength by a $NOON$ state projection measurement. Such
projection measurement can be generalized to an arbitrary state.
For the state from parametric down-conversion, its orthogonal
state projection measurement may lead to Heisenberg limit in phase
measurement.

\begin{acknowledgments}
This work was funded by the Chinese National Fundamental Research
Program (2001CB309300), the Innovation funds from Chinese Academy
of Sciences, and National Natural Science Foundation of China
(Grant Nos. 60121503 and 10404027). ZYO is also supported by the
US National Science Foundation under Grant No. 0427647.
\end{acknowledgments}

\end{document}